\documentclass[twocolumn,showpacs,preprintnumbers,amsmath,amssymb]{revtex4}
\usepackage{graphicx}% Include figure files
\usepackage{dcolumn}% Align table columns on decimal point
\usepackage{bm}% bold math
\usepackage{epstopdf}
\newcommand{\rr}{{\bf r}}
\newcommand{\xx}{{\bf x}}
\newcommand{\vv}{{\bf v}}

\newcommand{\FF}{{\bf F}}
\newcommand{\qq}{{\bf q}}
\newcommand{\nn}{{\bf n}}

\newcommand{\BE}{\begin{equation}}
\newcommand{\EE}{\end{equation}}
\newcommand{\be}{\begin{equation}}
\newcommand{\ee}{\end{equation}}
\begin{document}

\preprint{ }

\title{Domain Growth, Budding, and Fission in Phase Separating 
Self-Assembled Fluid Bilayers}
\author{Mohamed Laradji$^{1,3}$ and P. B. Sunil Kumar$^{2,3}$}
\affiliation {$^1$Department of Physics, The University of Memphis, 
Memphis, TN 38152 \\
$^2$Department of Physics, Indian Institute of Technology Madras,
Chennai 600036, India\\
$^3$MEMPHYS--Center for Biomembrane Physics, University of Southern Denmark, DK-5230, Denmark}

\date{\today}

\begin{abstract}

A systematic investigation of the phase separation dynamics in 
self-assembled multi-component bilayer fluid vesicles and 
open membranes is presented.
We use  large-scale dissipative particle dynamics to   
explicitly account for solvent, thereby allowing for 
numerical investigation
of the effects of hydrodynamics and area-to-volume constraints.
In the case of asymmetric lipid composition, we observed regimes corresponding
 to coalescence of flat patches, budding, vesiculation 
and coalescence of caps. The  area-to-volume constraint and hydrodynamics have
 a strong influence on these regimes and the crossovers between them. 
In the case of symmetric mixtures, irrespective of the area-to-volume ratio, 
we observed a growth regime with an exponent of $1/2$. The same exponent is also found in the 
case of open membranes with symmetric composition.
\end{abstract}


\maketitle
\section{INTRODUCTION}
Biomembranes are fascinating self-assembled quasi-two-dimensional 
complex fluids composed essentially of phospholipids and cholesterol.
The primary roles of biomembranes are 
the separation between the inner and outer environments of the cell 
or inner organelles, and
the support of an amazing specialized protein-based machinery which is crucial for a variety of
physiological functions, transmembrane transport and structural integrity of the cell~\cite{alberts94}. 
Many recent experiments demonstrated that biomembranes of eukariotic cells are laterally organized
into small nanoscopic domains, called rafts, which are rich in sphingomyelin and cholesterol. 
The higher content of cholesterol in rafts is due to the fact that the acyl chains in sphingomyelin are
mainly saturated, thereby promoting their interaction with cholesterol. 
Although, it is largely believed that this inplane organization is essential for a variety of physiological
functions such as signaling, recruitment of specific proteins and 
endocytosis~\cite{gousset02},  elucidation  of the fundamental
issues including the mechanisms leading to the formation of lipid rafts, their stability, 
and finite size remain elusive. 
Clearly, raft formation in biomembranes is complicated by the presence of 
many non-equilibrium mechanisms. In view of this, it is important to 
understand  the equilibrium phase behavior and the kinetics 
of fluid multicomponent lipid membranes before attempts are made to find
the effects of  more complex mechanisms that may be involved in the formation and 
stability of lipid rafts. 

The dynamics of in-plane demixing in multicomponent lipid membranes 
is richer than their counterparts in Euclidean three- or two-dimensional systems. This is largely due to 
(i) the strong coupling between the lipid composition and the membrane conformation, (ii) the
difference between the viscosities of the lipid bilayer and that of the embedding fluid, and (iii) the 
area-to-volume constraint, maintained by a gradient in osmotic pressure across the membrane. As a result,
various growth regimes, may be observed in multicomponent lipid membranes.
Phase separation dynamics in multicomponent vesicles following a quench from a single homogeneous phase to
the two-phase liquid-liquid coexistence region of the phase diagram has previously been considered by
means of a generalized time-dependent Ginzburg-Landau model on a non-Euclidean closed surface~\cite{taniguchi96,voth05}.
Limitations imposed by the parameterization of surface deformations  did not 
allow for budding  in this study. More recent simulations using a
dynamic triangulation Monte Carlo model~\cite{kumar96,kumar98,kumar01}, 
predicted dynamics which are much more complex than that in Euclidean surfaces. 
In particular, 
these simulations showed that for symmetric composition of binary mixtures,
 at intermediate times, the dynamics is characterized by the presence of
the familiar labyrinth-like spinodal pattern. At later times, in the presence of curvature 
composition coupling,  these patterns break up
leading to isolated islands. At  still later times, and
in the case of tensionless membranes, these domains reshape into buds connected to the parent membrane by very narrow
necks. Further domain growth proceeds via the Brownian diffusion of these buds, and their coalescence. 
It is important to remark that both the generalized time-dependent Ginzburg-Landau model and the Monte Carlo
dynamic triangulation model (1) do not account for the solvent, and are therefore purely dissipative,
(2) cannot account for the constraint on the volume enclosed by the vesicle, (3) conserve the topology of the
membrane throughout the simulation, and therefore do not account for fission or fusion processes. A model that
accounts of these effects is clearly warranted in order to compare with experiments. In order to achieve 
this we used a model based on the dissipative particle dynamics (DPD) approach
~\cite{laradji04}.

On the experimental side, recent  studies using advanced techniques such as 
two-photon fluorescence and confocal microscopy,
performed on giant unilamellar vesicles (GUVs) composed of 
dioleoylphosphatidylcholine,
sphingomyelin, and cholesterol,
showed the existence of lipid domains~\cite{dietrich01}.
More recent experiments on ternary mixtures composed of saturated  and
unsaturated phosphatidylcholine and cholesterol~\cite{silvius96,keller02,baumgart03}
saw the existence of liquid-liquid coexistence
over a wide range of compositions, an indication that liquid-liquid coexistence in lipid membranes is
ubiquitous to a wide variety of ternary lipid mixtures. It must be noted, however, that domains observed 
in these experiments are comparable to the size of the vesicle (micron-scale), orders of magnitudes larger
than rafts in biomembranes. Some of these experiments reported structures with many, more-or-less, curved
domains. But these are more akin to caps than fully developed buds. An important question that 
arises out of these studies is: (1) What role does the solvent hydrodynamics and the volume constraints play during coarsening
in multicomponent fluid vesicles?  (2) Does the  topology change drastically alter the kinetic pathway predicted by simulations that does not allow for topology changes?

In this paper, we present results from extensive numerical simulations of
self-assembled lipid vesicles and open membranes using the dissipative particle 
dynamics approach. We specifically investigated the phase separation dynamics
in lipid membranes following a quench from the one phase region to the 
fluid-fluid coexistence region of the phase diagram.
The lipid membrane is composed of self-assembled lipid particles in an explicit solvent, 
thus accounting for hydrodynamic effects. Furthermore, the parameters of the model 
are such that the membrane is impermeable to the solvent,
thus allowing us to investigate the effect of area-to-volume ratio on the 
dynamics. We specifically investigated the effects of (1) area-to-volume ratio, 
(2) line tension, and (3) lipids composition on the dynamics. We find that the
path,through which dynamics proceeds, depends on the area-to-volume ratio 
and composition. In off-critical quenches, in particular, the dynamics proceeds 
via the coalescence of small flat patches at intermediate times, followed 
by their budding and vesiculation. At late times, domain growth proceeds via 
coalescence of caps remaining on the vesicle. Crossovers between these 
regimes are strongly affected by the area-to-volume ratio and line tension. 
In the case of critical quenches, domain growth proceeds via dynamics similar 
to that in Euclidian two-dimensional fluids. That is, in this case the effect on 
the embedding fluid on the coarsening process seems to be not very obvious.
We check this by also by simulating critical quenches in open membranes with different projected areas.

This article is organized as follows: in Sec. II, the model and simulation 
technique are presented.  In Sec. III, results of our simulations are presented. 
Finally, we summarize and conclude in Sec. IV.

\section{MODEL AND METHOD}

In this work, the dynamics of phase de-mixing of a two-component lipid mixture in an explicit 
solvent is investigated using dissipative particle dynamics (DPD) approach. DPD was first 
introduced by Hoogerbrugge and Koelman more than a decade ago~\cite{hoogerbrugge92}, and was cast 
in its present form about five years later~\cite{espagnol95,espagnol97}.  DPD is reminiscent 
of molecular dynamics, but is more appropriate for the investigation of generic properties of 
macromolecular systems. The use of soft repulsive interactions in DPD, allows for larger 
integration time increments than in a typical molecular dynamics simulation using 
Lennard-Jones interactions. Thus time and length scales much larger than those in 
atomistic molecular dynamics simulations can be probed by the DPD approach. 
Furthermore, DPD uses pairwise random and dissipative forces between neighboring 
particles, which are interrelated through the fluctuation-dissipation theorem. 
The pairwise nature of these forces ensures local conservation of momentum, a necessary
condition for correct long-range hydrodynamics~\cite{pagonabarraga98}.

The system is composed of simple solvent particles (denoted as $w$), and two types 
of complex lipid particles (A and B).  A lipid particle is modeled as a flexible 
amphiphilic chain with one hydrophilic particle, mimicking a lipid head
group, and three hydrophobic particles, mimicking the lipid acyl-group. 
More complicated lipid structures and artifacts due to the choice of 
 simulation parameters have been investigated by other groups~\cite{ask05,shillcock02,ilya05}. 
These details are not expected to affect the qualitative nature of the results 
reported here.

\begin{figure}
\includegraphics[scale=0.3]{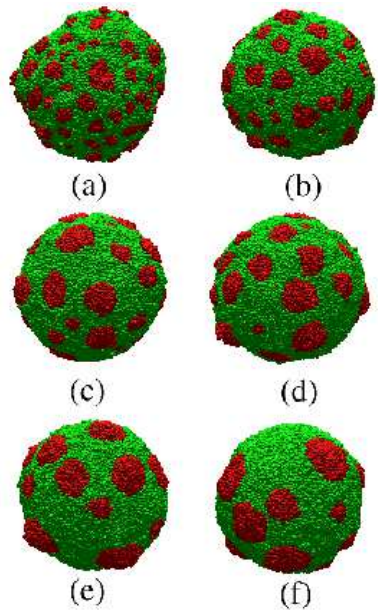}
\caption{Snapshots of a phase separating vesicle with low excess area,
and $\phi_B=0.5$ at (a) $t=100\,\tau$, (b) $500\,\tau$, (c) $1000\, \tau$,
(d) $2000\, \tau$, (e) $3000\, \tau$, and (f) $4000\, \tau$.}
\label{caseIsnap}
\end{figure}

The heads 
of the A and B lipids are denoted by $h_A$ and $h_B$, and their respective tails 
are denoted by $t_A$ and $t_B$.  For simplicity, we focus in this study on the case 
where the interactions are symmetric under the exchange between $A$ and $B$ lipids. 
Thus this model does not contain any explicit coupling between curvature and composition. 
The time evolution of the position and velocity of each dpd particle $i$, denoted by $({\bf r}_i,{\bf v}_i)$, 
are governed by Hamilton's equations of motion. The three pairwise forces are given by
\begin{eqnarray}
\FF^{(C)}_{ij}&=& a_{ij}\omega(r_{ij})\nn_{ij}, \label{eq:conservative-force}\\
\FF^{(D)}_{ij}&=& \gamma_{ij}\omega^2(r_{ij})(\hat\rr_{ij}\cdot\vv_{ij})\nn_{ij}, 
\label{eq:dissipative-force}\\
\FF^{(R)}_{ij}&=& \sigma_{ij}(\Delta t)^{1/2}\omega(r_{ij})\theta_{ij}\nn_{ij}, 
\label{eq:random-force}
\end{eqnarray}
where $r_{ij}=\rr_j-\rr_i$, ${\nn}_{ij}={\rr_{ij}}/{r_{ij}}$, and $\vv_{ij}=\vv_j-\vv_i$.
$\theta_{ij}$ is a symmetric random variable satisfying
\begin{eqnarray}\label{theta_equations}
\langle\theta_{ij}(t)\rangle&=&0,\\
\langle\theta_{ij}(t)\theta_{kl}(t')\rangle
&=&(\delta_{ik}\delta_{jl}+\delta_{il}\delta_{jk})\delta(t-t').
\end{eqnarray}
with $i\ne j$ and $k\ne l$. In Eq.(\ref{eq:random-force}), $\Delta t$ is the iteration time step. The weight factor is
chosen as 
\BE\label{eq:omega}
\omega(r)=\left\{\begin{array}{ll}
1-r/r_c & \mbox{ for $r \leq r_c$,}\\
0 & \mbox{ for $r > r_c$}
\end{array}
\right.
\EE
where $r_c$ is the interactions cutoff radius. The choice of $\omega$ in Eq. (\ref{eq:omega}) 
ensures that the interactions are all soft and repulsive. The equations of motion of 
particle $i$ are given by 
\begin{eqnarray}\label{eq:motion}
d\rr_i(t)&=&\vv_i(t) dt,\\
d\vv_i(t)&=&\frac{1}{m}\biggl{(}\sum_j\FF^{(C)}_{ij} dt+\sum_j\FF^{(D)}_{ij} dt\nonumber\\
& & + \sum_j\FF_{ij}^{(R)} (dt)^{1/2}\biggl{)},
\end{eqnarray}
where $m$ is the mass of a single DPD particle. Here, for simplicity, masses of all 
types of dpd particles are supposed to be equal.
Assuming that the system is in a heat bath at a temperature $T$, the  
parameters $\gamma_{ij}$ and $\sigma_{ij}$ in Eqs.~(\ref{eq:dissipative-force} 
and \ref{eq:random-force}) are related to each other by the fluctuation-dissipation theorem, 
\BE \label{eq:fluctuation-dissipation}
\gamma_{ij}=\sigma_{ij}^2/2k_{\rm B}T.
\EE
The parameters, $a_{ij}$, of the conservative forces are specifically chosen as
\BE
a_{ij}=\frac{\epsilon}{r_c} \left(\begin{array}{cccccc}
   &h_A &t_A  &w  &h_B  &t_B\\
h_A&25  &200  &25 &a_{AB}  &200\\
t_A&200 &25   &200&200  &a_{AB}\\
w  &25  &200  &25 &25   &200\\
h_B&a_{AB} &200  &25 &25   &200\\
t_B&200 &a_{AB}  &200&200  &25
\end{array}
\right),
\EE
where $\epsilon$ and the cutoff radius, $r_c$, set the energy and length scales, respectively.
The effect of line tension is studied by varying the parameter $a_{AB}$.
The integrity of a
lipid particle is ensured via an additional simple harmonic interaction, between consecutive particles, whose force
is given by
\BE\label{eq:spring-force}
{\bf F}_{i,i+1}^{(S)}=-C \left(1-r_{i,i+1}/b\right){\nn}_{i,i+1},
\EE
where we set, for the spring constant and the preferred bond length, 
values $C=100\epsilon$ and $b=0.45r_c$, respectively.

\begin{figure}
\includegraphics[scale=1.0]{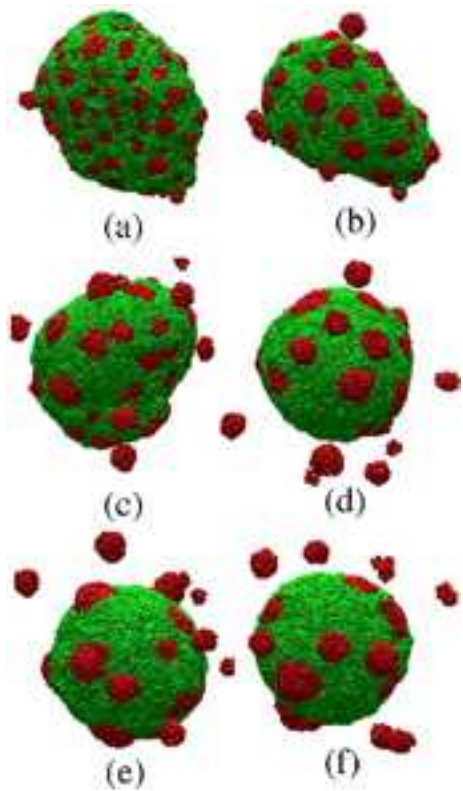}
\caption{Snapshots of a phase separating vesicle in case-II. The time sequence
of the snapshots is the same as in Fig.~1.}
\end{figure}

In our simulations, we used $\sigma=3.0\left(\epsilon^3 m /r_c^2\right)^{1/4}$. Most of the simulations were performed
at $k_BT=\epsilon$, and a fluid density $\rho=3.0r_c^{-3}$. The iteration time was chosen 
to be $\Delta t=0.05 \tau$~\cite{footnote1}, with 
the time scale $\tau=(mr_c^2/\epsilon)^{1/2}$. The equations of motion are integrated using velocity-Verlet
algorithm~\cite{gerhard}. The total number of lipid particles used was $16\ 000$, and both 
cases of closed vesicles and open membranes were simulated. In the case of a closed vesicle, the box size
is $(80\times 80\times 80) r_c^3$ corresponding to a total number of $1\ 536\ 000$ dpd-particle. 
In the case of open membranes, we consider box sizes of $L_x=L_y> L_z$, with $L_z=40r_c$, such that the fluid
density is equal to that in the case of closed vesicles.
Open membranes with different tensions are simulated by varying $L_x$, such that the number of lipid particles 
and fluid density are kept constant. Note that $L_z=40 r_c$ is still much larger than the
thickness of the bilayer. Periodic boundary conditions are applied in all directions for 
both cases of closed and open membranes.

\begin{figure}
\includegraphics[scale=0.5]{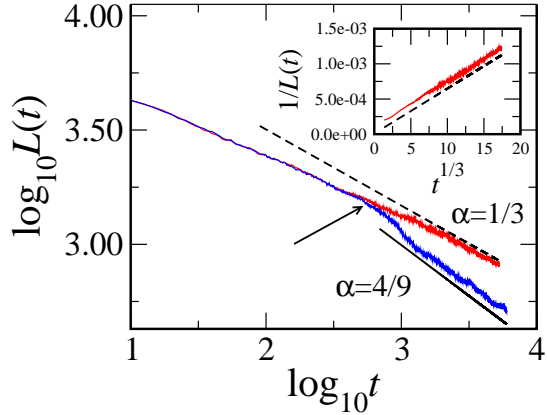}
\caption{Net interface length as a function of time in case-I (blue curve) and
case II (red curve). The green curve corresponds to case II with low line
tension (see text for details of corresponding interactions). 
The dashed lines and the dot-dashed line
have slopes of $0.3$ and $4/9$, respectively. 
In the inset, the number of vesiculated domains is shown as a function of 
time for case-II with high line tension (corresponding to the red curve).
The arrows point to the time regime during which budding and 
vesiculation occur in case II with high line tension.}
\label{interface}
\end{figure}

\begin{figure}
\includegraphics[scale=0.5]{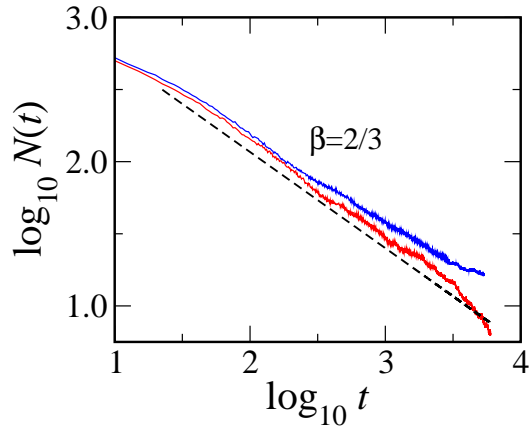}
\caption{Number of clusters on the vesicle as a function of time. Curve colors 
correspond to those in Fig.~\ref{interface}. The slope of the dashed line 
is 2/3. Note that in the case of high excess area, clusters that have vesiculated are
excluded.} 
\label{cluster}
\end{figure}

Previous simulations based on DPD models have shown that open bilayers and closed vesicles can
be self-assembled~\cite{shillcock02,yamamoto02,yamamoto03}. These studies, however, 
were performed on smaller systems than those in the present study. In order to save 
computer time, we prepare our vesicles starting from an almost closed 
configuration, composed from a single type of lipid. This approach allows for 
the equilibration of the lipid surface 
coverage in both leaflets through the diffusion of the lipids via 
the rim of the open vesicle. We find that the vesicle
typically closes within about $50\ 000$ DPD time steps. 
Once the vesicle is closed, we also found that within 
our parameters, the number of solvent
particles inside the vesicle and the numbers of lipid particles in the 
inner and outer leaflets remain constant throughout
the simulation. A vesicle composed of $16\ 000$ lipid chains, 
prepared as indicated above, is found to be nearly 
spherical and contain about $138\ 500$ solvent particles inside it. 
Vesicles with excess area (high area-to-volume ratio) are
then prepared by transferring solvent particles from the core of 
the vesicles, prepared as indicated above, to the outer 
region, such that the fluid density is kept constant. 
An open membrane is prepared by placing a bilayer parallel to the $xy$-plane at $z=L_z/2$.
The membrane is let to equilibrate until fluctuations of its height attain saturation.
After equilibration of the closed vesicle or the open membrane, 
the phase separation process is triggered through
an instantaneous change of a fraction of the A-lipids to B-lipids 
such that their composition is equal to $\phi_B$. This
mimics a quench from a homogeneous state to the two-phase region. 

\begin{figure}
\includegraphics[scale=0.5]{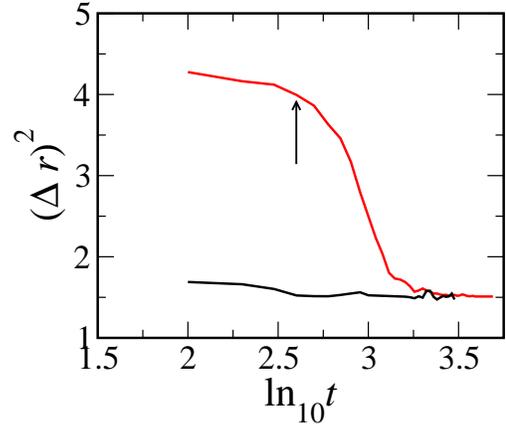}
\caption{Mean square of the distance of the A-lipids from the center of mass. 
The top and bottom curves correspond to $\nu=0.17$ and $0.14$, respectively.} 
\label{meansquare}
\end{figure}

We have performed systematic simulations in which 
following parameters were varied:\\
(i) The strength of the repulsive interaction between A and B lipids, $a_{AB}$, 
in order to infer the effect of line tension, $\lambda$, between A and B domains. 
This parameter was varied for both open and closed membranes.\\
(ii) The compositions of the two lipids, $\phi_A$ and $\phi_B=1-\phi_A$. We considered the cases of $\phi_A=0.5$ 
and $0.3$. This parameter was varied for both closed and open membranes. \\
(iii) The area-to-volume ratio, $\nu$, in the case of closed vesicles, defined here as 
$\nu=\left(N_{head}+N_{tail}\right)/{N_w}$, where $N_{head}$, $N_{tail}$ are the total numbers of head and tail 
dpd particles, respectively, and $N_w$ is the total number of solvent particles inside the vesicle.
In the case of open membranes, the projected area, effectively plays the role of area-to-volume 
ratio in closed vesicle.

In the presentation and discussion of our results, we use the following labels to indicate the 
parameters used for different simulated systems

\begin{center}
\begin{tabular}{|c|c|c|c|}\hline\hline
\ \ \ {\em system}\ \ \  &\ \ \ \ \ \  $\nu$\ \ \ \ \ \  &\ \ \  $a_{AB}\ \ \ $ &\ \ \  $\phi_B$\ \ \ \\ \hline\hline
${\cal A}^{(100)}_1$ & $0.462$ & $100$   & $0.3$ \\ \hline
${\cal A}^{(100)}_2$ & $0.567$ & $100$   & $0.3$ \\ \hline
${\cal A}^{(86)}_2$ & $0.567$ & $86$     & $0.3$ \\ \hline
${\cal A}^{(68)}_2$ & $0.567$ & $68$     & $0.3$ \\ \hline
${\cal A}^{(50)}_2$ & $0.567$ & $50$     & $0.3$ \\ \hline
${\cal C}^{(100)}_1$ & $0.462$ & $100$   & $0.5$ \\ \hline
${\cal C}^{(100)}_2$ & $0.567$ & $50$    & $0.5$ \\ \hline
${\cal C}^{(50)}_2$ & $0.567$ & $50$     & $0.5$ \\ \hline\hline
\end{tabular}
\end{center}

The interaction parameters of the present model are selected such that 
the membrane is impermeable to solvent particles. 
This implies that the number of solvent particles inside closed vesicles 
is constant, thereby allowing us to effectively 
investigate the effect of area-to-volume ratio.
In experiments, this parameter is controlled via the osmotic pressure.

\begin{figure}
\includegraphics[scale=1.0]{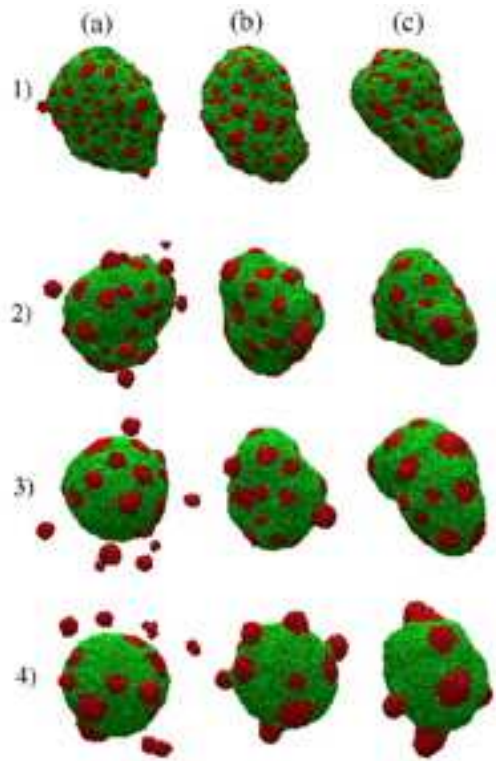}
\caption{Snapshot sequences for closed vesicles with excess area parameter $\nu=0.17$.
(a), (b) and (c) correspond to $a_{AB}=100\epsilon$, $68\epsilon$
and $50\epsilon$, respectively. Snapshots from top to bottom correspond to $t=100\tau$, $1000\tau$,
$2000\tau$, and $4000\tau$, respectively.}
\label{lowtensIIsnaps}
\end{figure}

\section{RESULTS AND DISCUSSION}

For the parameter values mentioned above, our model membrane does not show any flip-flop 
motion of the lipids, {\it i.e.} throughout the simulation time the number of lipids in the 
upper and lower membrane  are the same.  We also find that the coupling between the compositions of the two leaflets 
is found to be very strong. This is not surprising, considering the fact that we have chosen 
$a_{t_At_B}>a_{t_At_A}=a_{t_Bt_B}$.

\noindent
\subsection{Case of closed vesicles with $\phi_B=0.3$}

In Fig.~1, snapshots of closed vesicles configurations 
in the case of $\phi_B=0.3$ and with small
area-to-volume ratio, corresponding to system ${\cal A}_1^{(100)}$ are shown. 
This figure shows that the phase separation process after a quench to the two-phase 
region proceeds in a manner similar to that in Euclidian systems, {\it i.e.}, 
through the formation of small domains, and their coalescence in time. During the 
early stages of the dynamics, domains have average curvatures that are equal to the 
surrounding majority component, an indication that during the early stages of the 
phase separation process, the curvature is decoupled from the
composition. This is due to the fact that the compositions in 
the two leaflets are equal, and the {\em A-} and {em B-}lipids
have identical architectures, leading to a decoupling between 
curvature and composition. As time evolves, the interface tension starts to 
assert and the  domains curve
very slightly from the majority component, while the vesicle becomes 
more spherical, an implication that an increase in tension has occurred.  
In Fig.~2, snapshots corresponding to $\phi_B=0.3$, but with a high 
excess area parameter , ${\cal A}_2^{(100)}$, are shown. The high excess area is obtained 
by removing  $25610$ solvent particles from the core of the vesicle
and putting them in the outer region, such that the fluid density 
is maintained.  Notice, here again, that during the early stages, 
domains have same curvature as the majority component, implying the 
decoupling between curvature and composition during these stages. 
At about  $400\tau$, B-domains start to curve away from the average vesicle's
curvature.

\begin{figure}
\includegraphics[scale=0.5]{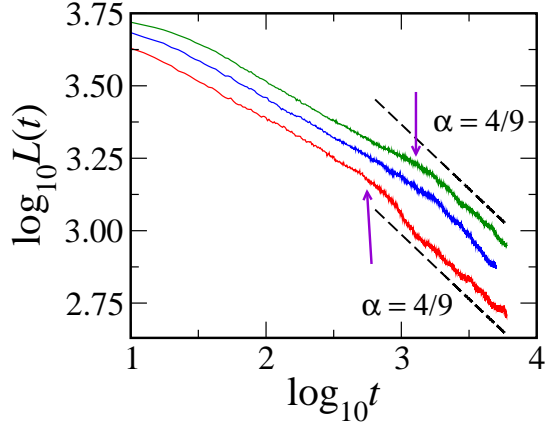}
\caption{Net interface length versus time for the case of closed vesicles with excess area parameter,
$\nu=0.17$, for different values of line tension.
Curves from top to bottom correspond to $a_{AB}=50\epsilon$, $68\epsilon$ and $100\epsilon$, respectively.}
\label{lowtensinterface}
\end{figure}

Domain growth is monitored through both the net interface length, 
$L(t)$, and the average domain size, $R(t)$, as calculated from the cluster 
size distribution. In order to see how the interfacial length 
can be used as a measure of domain coarsening, consider a closed vesicle
formed of $N (t)$ B-domains at some time $t$, with an average linear 
size $R$. Furthermore, let us consider the case where
the domains are circular. The net interface length is
then given by $L(t)=2\pi N(t) R(t)$. Since the net amount 
of the B-component, is conserved, we also have for the area
occupied by the B-component, ${\cal A}_B=\pi N(t) R^2(t)$. We therefore have 
\BE 
L(t)\sim {\cal A}_B/R(t),
\EE
and the number of B-domains is therefore given by
\BE\label{eq:intlength_radius}
N(t)\sim{\cal A}_B/R^2(t).
\EE
In Fig.~3, $L(t)$ versus time is shown for the case of $\phi_A=0.3$ 
and for low and high area-to-volume parameter. We notice from this figure 
that, during intermediate times, {\it i.e.} $t<400\tau$, the interfacial 
length is independent of the area-to-volume ratio, and has the form 
\BE
L(t)\sim t^{-\alpha},
\EE
with the growth exponent, $\alpha \approx 0.3$.
The number of domains as calculated from the cluster size distribution 
is shown in Fig.~4, which again shows that the
number of clusters is independent of the area-to-volume ratio at intermediate times, and that 
\BE
N(t)\sim t^{-\beta},
\EE
with $\beta\approx 2\alpha\approx 2/3$, in agreement 
with Eq.~(\ref{eq:intlength_radius}).

\begin{figure}
\includegraphics[scale=0.5]{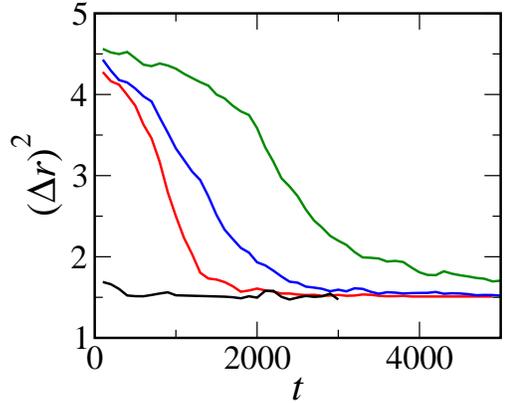}
\caption{The mean square of the distance of the A-lipid heads from the center 
of mass, $(\Delta r)^2$ versus time. Curves from top to bottom correspond to
systems ${\cal A}_2^{(50)}$, ${\cal A}_2^{(68)}$, ${\cal A}_2^{(100)}$, and 
${\cal A}_1^{(100)}$.}
\label{meansquareII}
\end{figure}

A growth exponent, $\alpha=1/3$, in Euclidian multicomponent 
systems is usually attributed to the evaporation-condensation mechanism as 
explained by the classical theory of Lifshitz and Slyozov
~\cite{lifshitz62}. In a fluid system such as ours, domain growth can also 
be the result of the motion of the entire domains themselves resulting in collision 
between domains leading to coalescence.

Two domains coalesce if they travel a distance, $l(t)$ determined by the average 
area on the membrane occupied by a domain. This is  given by 
\BE 
l(t)^2\sim {\cal A}/N(t).
\EE
Moreover, assuming
that the domains perform a Brownian walk before their collision, 
the collision time should obey $l(t)^2\sim D_R t$. In the 
case of an isolated  two-dimensional 
fluid, as can be calculated from the  Stoke's formula for the 
drag on a circular domain due to the surrounding fluid,  
the diffusion coefficient is independent of 
the domain size. However, in our case, the drag experienced 
by the domains results mainly from the three-dimensional embedding 
fluid, leading to $D_R\sim 1/R$. Using this fact, we have the time 
dependence of domain size as 

\begin{figure}
\includegraphics[scale=0.5]{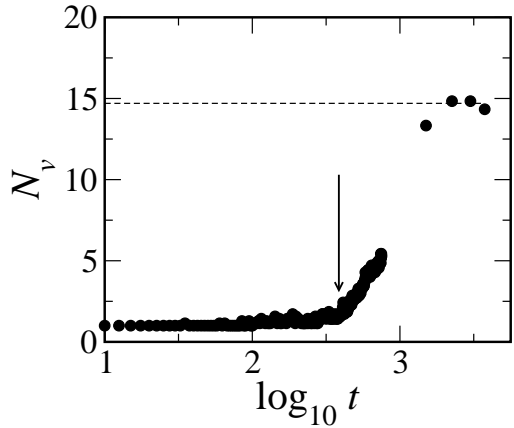}
\caption{The number of vesiculated buds versus time for system ${\cal A}_2^{(100)}$.}
\label{numbervesbud}
\end{figure}

\BE
R(t)\sim t^{1/3},
\EE
and the number of clusters
\BE
N(t)\sim t^{-2/3},
\EE
in good agreement with our numerical results.

Irrespective of the initial excess area, the very late times configuration
is always that of a tensed vesicle. This can be seen from the mean square of
of the positions of  the A-lipid heads from the center of mass of the vesicle,
$(\Delta r)^2$.
This is shown in Fig.~5 for the case of $a_{AB}=100\epsilon$, for systems 
 ${\cal A}_1^{(100)}$ and ${\cal A}_2^{(100)}$. A large $(\Delta r)^2$ implies a floppy vesicle with a lot of 
excess area. Notice that $(\Delta r)^2$ decreases rapidly 
after about $400\tau$, eventually reaching a value equal to that without 
excess area.  The fast decrease in $(\Delta r)^2$ 
is due to the reshaping of the domains into caps, the budding  
of some of them and their vesiculation. The buds, 
once formed, are found to vesiculate within a short period of time, 
of order $10\tau$. We confirm this by performing simulations of 
a single B-domain occupying $12\%$ of the total area of a vesicle 
with excess area.  The fission mechanism of a bud itself  is 
a very interesting phenomena, and has recently been investigated in
detail using DPD simulations~\cite{yamamoto02}.

\begin{figure}
\includegraphics[scale=1.0]{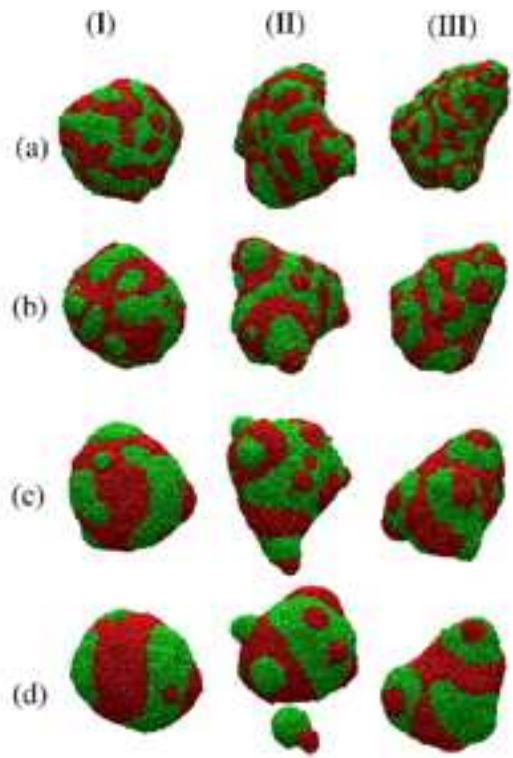}
\caption{Snapshot sequences of closed vesicles with $\phi_B=0.5$.
(I), (II) and (III) correspond to systems ${\cal C}_1^{(100)}$, ${\cal C}_2^{(100)}$ and
${\cal C}_2^{(50)}$, respectively. Snapshots from top to bottom correspond to
$t=100\tau$, $200\tau$, $400\tau$ and $700\tau$, respectively.}
\label{symvessnaps}
\end{figure}

As a result of vesiculation, the parent vesicle looses most of its excess 
area, leading it to acquire a more spherical shape, as shown in Fig.~2,
and implied by Fig.~5.  This induced lateral tension prevents the 
domains from further capping. Further domain coarsening may therefore 
proceed via the coalescence of these caps. 

\subsection{Effect of line tension on the phase separation in closed 
vesicles with $\phi_B=0.3$}

The effect of line tension on the dynamics of domain growth in closed 
vesicles with $\phi_B=0.3$ is investigated by
performing simulations at different values of $a_{AB} 
=100\epsilon$, $86\epsilon$, $68\epsilon$, 
and $50\epsilon$ in the presence  of large excess area. 

In Fig.~6, snapshots for the cases of $a_{AB}=100\epsilon$, 
$68\epsilon$ and  $50\epsilon$ are shown for comparison. These snapshots 
clearly show that the line tension play an important role on the dynamics. 
Corresponding interfacial lengths as a function of 
time are shown in Fig.~7.  This figure again shows that budding is 
delayed as the line tension is increased. 
We also notice that while in the system with $a_{AB}=100\epsilon$, 
budding of domains is followed by their vesiculation,
in the systems with $a_{AB}=86\epsilon$ and $68\epsilon$,
 very few buds vesiculate. No vesiculation was observed in 
the case of low line tension ($a_{AB}=50\epsilon$). 

In order to gain further insight into the effect of line tension 
on domains caping, let us consider a circular domain of  B-phase 
with area $a$, surrounded by a sea of A-phase, on a tensionless 
membrane. Let $c$ be the absolute value of the
mean curvature of this domain which is assumed to be uniform.
The free energy of the domain is 
therefore given by
\BE
{\cal E}_a=2{\kappa}a c^2+{\lambda}{l},
\EE
where $\kappa$ and $\lambda$ are the bending modulus and line
 tension, respectively, and $l$ is the perimeter of
the domains, given by
\BE
{l}=2\pi\left(\frac{a}{\pi}\right)^{1/2}\left(1-\frac{ac^2}{4\pi}\right)^{1/2}.
\EE
The free energy is then rewritten as 
\BE\label{eq:free-energy}
{\cal E}=8\pi\kappa\left[{\tilde c}^2+\frac{\lambda}{2\kappa c_{max}^2}\left(1-{\tilde c}^2\right)^{1/2}\right],
\EE
where $a=4\pi/c_{max}^2$ and ${\tilde c}=c/c_{max}$. The free 
energy in Eq.~(\ref{eq:free-energy}) has a minimum
at $c=0$. This minimum is absolute if the area of the domain is smaller than $a_0=4\pi(\kappa/\lambda)^2$.
Otherwise, the free energy is lower for $c> 0$. 
These calculations imply that for a given $\kappa/\lambda$, the
onset of domain capping occurs when their average radius exceeds $R_0=2\kappa/\lambda$, in qualitative agreement with our
simulation results. 

\begin{figure}
\includegraphics[scale=0.5]{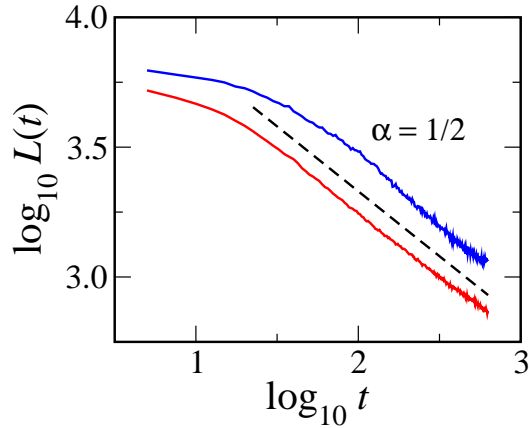}
\caption{Net interface length versus time for the case of closed vesicles with $\phi_B=0.5$. Top and bottom curves
correspond to  ${\cal C}_2^{(100)}$ and ${\cal C}_2^{(50)}$, respectively. The slope of the dotted line is 0.5.}
\label{syminterface}
\end{figure}

In order to verify the arguments above, 
the bending modulus and line tension for the cases of $a_{AB}=50$ and 
$100\epsilon$ are extracted numerically. Details of the numerical approach used to derive these quantities are
described in Appendices A and B. We obtain a bending modulus $\kappa\approx 8.4\epsilon$,
 and a line tension $\lambda= 7.1\epsilon/r_c$ and $5.4\epsilon/r_c$.  The number 
of domains at the onset of capping is calculated as
\BE\label{eq:nbr-domains}
N_c=\frac{{\cal A}_v\phi_B}{\pi R_0^2},
\EE
where ${\cal A}_v$ is the vesicles area. In the following table, the 
number of domains at the onset of capping, $N_c$, obtained 
using Eq.~(\ref{eq:nbr-domains}) and  that obtained from the simulations
are shown.

\begin{center}
\begin{tabular}{|c|c|c|}\hline\hline
\ \ \ \ \ \  &\ \ \ \ \ \  $A^{100}_2  $\ \ \ \ \ \  &\ \ \  $A^{(50)}_2$\ \ \ \ \ \\ \hline\hline
$N_c$ (from Eq. (\ref{eq:nbr-domains})) & 91          & 54 \\   \hline
$N_c$ (from simulation)                & 71          &  38 \\ \hline\hline
\end{tabular}
\end{center}
It is interesting to note that our numerical results are fairly in good 
agreement with the theoretical values. The discrepancy  is 
reasonable considering the fact that in Eq.~(\ref{eq:free-energy}), 
only lowest order terms, in curvature, are accounted for. In the simulation, however, a cap is not uniformly curved.

Since domain growth prior to capping is governed by a $t^{1/3}$ law, 
the onset time of domain capping should be, $t_{cr}\sim (\kappa/\lambda)^3$. 
In the simulation, this time scale is found to be approximately $400 \tau$ and 
$1000\tau$ for systems $A^{(100)}_2$ and
$A^{(50)}_2$, respectively. The ratio between these two times is very 
close to the ratio of line tensions given by  
$\left(\lambda_{A^{(50)}_2}/\lambda_{A^{(100)}_2}\right)^3\approx 0.45$. 

\begin{figure}
\includegraphics[scale=1.0]{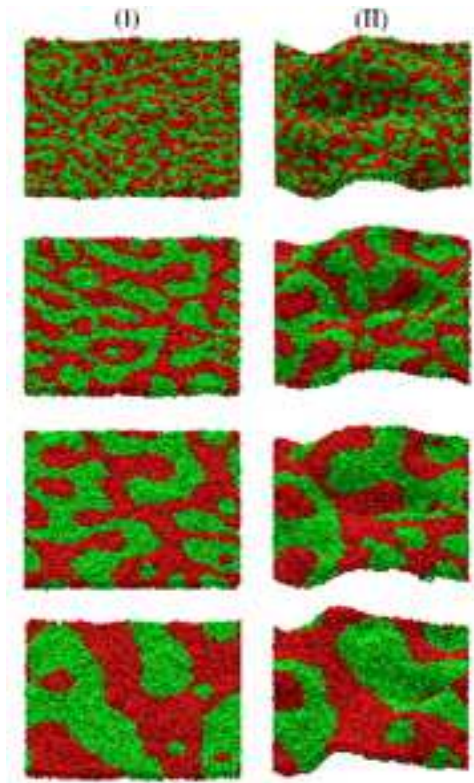}
\caption{Snapshot sequences for open membranes with $\phi_B=0.5$, with $a_{AB}=50$.
(I) and (II) correspond to projected areas $L_x\times L_y=86r_c\times 86r_c$ and $78r_c\times 78r_c$,
respectively. Snapshots from top to bottom correspond to
$t=50\tau$, $200\tau$, $400\tau$ and $800\tau$, respectively.}
\label{symopensnaps}
\end{figure}

In Fig.~8, the mean square of the distance of A-lipid heads from the vesicle's center of mass, $(\Delta r)^2$,
is plotted for systems, ${\cal A}_2^{(100)}$, ${\cal A}_2^{(68)}$, ${\cal A}_2^{(50)}$, 
together with the case with small excess area ${\cal A}_1^{(100)}$. This figure clearly illustrates that domains capping 
is shifted towards later times as the line tension is decreased.

\subsection{Late time dynamics of closed vesicles with excess area and $\phi_B=0.3$}

The dynamics in systems ${\cal A}^{(100)}_1$ and ${\cal A}^{(100)}_2$ departs from each other after about $t=400\tau$,
as shown in Fig.~3: The dynamics of domain growth in the case of high excess area speeds up at
late times, as compared to the case with low excess area.
As shown in Fig.~3, most domains of the minority component in the case with low excess area
remain flat with a curvature equal to that of the majority
component. Capping, in this case, is suppressed due to the lateral tension induced by volume constraint and low excess area.
In contrast, in the case where excess area is high, domains reshape into caps, allowing the interfacial length to
decrease rapidly.
The presence of excess area allows some of the caps to further reshape into buds as shown in Fig.~2, which then vesiculate, since the line tension is relatively high in this case.
Bud vesiculation occurs over a relatively short period of time, as indicated in Fig.~9.  We confirm this by performing
simulations of a vesicle composed of two coexisting domains and with high excess area. We found that once the bud is formed,
it vesiculates within a time period of about $10\tau$.

\begin{figure}
\includegraphics[scale=0.5]{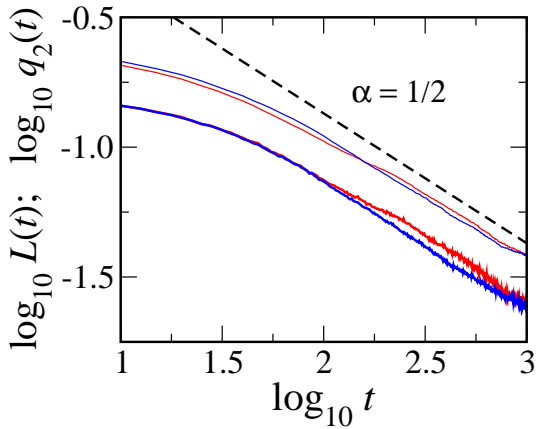}
\caption{Interfacial length (bold curves) 
and the square root of the second moment (thin curves) versus 
time for open membranes with $\phi_B=0.5$, with $a_{AB}=50$. Red and blue
curves correspond to a projected area $L_x\times L_y=86r_c\times 86 r_c$ and
$78r_c\times 78 r_c$, respectively. The slope of the dotted line is $-1/2$.} 
\label{openmeansquare}
\end{figure}

The vesiculation of some B-domains results in a marked decrease of
excess area leading the vesicle with the remaining
B-domains to acquire a much more spherical shape.
Once all excess area has been released, domain growth crosses over to a regime
characterized by $L(t)\sim t^{-\alpha}$, 
with $\alpha\approx 0.4$, as shown in Fig.~3. Fig.~4 demonstrates that
the number of domains remaining on the
vesicle decreases with time as $N_B\sim t^{-\beta}$ with a 
growth exponent $\beta\approx 2/3$. 

We believe that the higher  growth exponent at late times, $\alpha\approx 0.4$, 
is the result of coalescence of well formed caps following their Brownian 
motion. Indeed, as Fig.~1 shows domains of the minority component are more 
akin to hemispherical caps at late times.  Assuming that the curvature of 
this caps are determined by the competition between the bending energy and 
the line tension at the interface between A and B lipids, we can write 
the average interfacial length, $l_c$, of a single cap as 

\BE
l_c\sim \left(\frac{\kappa a_c}{\lambda}\right)^{1/3}.
\EE

Domain growth proceeds via the Brownian motion of domains, leading to 
their coalescence as they collide. Therefore, the mean
square of the distance travelled by a single domain,
\BE
d^2\sim Dt,
\EE
where $D\sim a_c^{-1/2}$, and $d^2\sim {\cal A}_v/N_c$, ${\cal A}_v$ 
being the area of the vesicle. Using as well, the fact 
that the net area of the $B$-domains is a constant of time, 
i.e., $N_c l_c^3\sim const$, one then obtains 
\BE
N_c\sim \frac{1}{t^{2/3}},
\EE
and the net interfacial length, $L_c=N_c l_c$ is then given by
\BE
L_c \sim t^{-4/9},
\EE
in excellent agreement with our numerical results.

\subsection{Case of closed and open membranes with $\phi_B=0.5$}

We now focus on the dynamics of phase separation of multicomponent closed 
 vesicles with symmetric volume fractions of the two components. In Fig.~10, 
snapshots corresponding to the case with high excess area and with high line 
tension are shown.
The corresponding interface length versus time is shown in Fig.~11. 
This figure shows that the characteristic domain size 
in the case of $\phi_B=0.5$ is much more pronounced than that with asymmetric 
composition. Furthermore the growth exponent at intermediate times is 
$\alpha\approx 1/2$, larger than that for the case of $\phi_B=0.3$. 

We must note that in the case of $\phi_B=0.3$, where the lipid
domain structure is circular, two measures where used to characterize domain 
growth. These correspond to the interface length and the average cluster size.
In the case of $\phi_B=0.5$, only the interface length was so far presented. 
In order to investigate the robustness of the corresponding
growth exponent, $\alpha=1/2$, we also performed simulations of open 
membranes extending along the $xy$-plane. These simulations allow us to extract
another length scale from the composition structure factor, 
$S(q,t)=\langle|\phi_{\bf q}(t)|^2\rangle/L_xL_y$, as $R(t)=2\pi/q_2$, where
\BE
q_2(t)=\left(\frac{\int dq q^2S(q,t)}{\int dq S(q,t)}\right)^{1/2}
\EE 
The effect of tension, and thus the influence of bending modes on phase 
separation, can also be investigated through varying the projected area, 
$L_xL_y$. 

In Fig.~12, sequences of snapshots of open membranes with $\phi_B=0.5$ and projected 
areas $L_xL_y=(80\times 80)r_c^2 $  and $(78 \times 78)r_c^2$ are displayed. Corresponding interface length, $L(t)$,
and the square root of the second moment, $q_2(t)$, are presented in Fig.~13. It is clear from
this figure that both lengths scale as $t^\alpha$ with $\alpha\approx 1/2$.

The exact mechanism leading to this exponent is not clear at present. A growth 
exponent, $\alpha=1$, has been predicted in the case of two component fluids with interconnected structures  in three-dimensions and is the result of the instability of the 
tubular domains against the peristaltic modes~\cite{siggia79}. 
Such instability does not exist in pure two-dimensional fluids
~\cite{sanmiguel85}. On the other hand, 
several simulations on purely two-dimensional 
fluids based on molecular dynamics~\cite{toxvaerd93}, model H~\cite{lookman95} 
and lattice gas~\cite{sunil03}, have seen a growth exponent, $\alpha=1/2$: 
A theoretical argument for this exponent is however lacking.
The lipid bilayer, being a two-dimensional fluid embedded in a three-dimensional
solvent, is clearly more complicated than a purely two- or three-dimensional 
fluid.

\section{Summary and Conclusions}

In summary, we presented a detailed study of the phase separation
dynamics of self-assembled bilayer fluid membranes, with hydrodynamic effects,
using dissipative particle dynamics. We considered both open and closed
membranes, and investigated the effect of composition, line tension and
surface tension. In all cases, hydrodynamics is found to affect the 
coarsening dynamics at all time scales.
In the case of closed vesicles with off-critical quenches, rich dynamics 
was observed, with crossovers depending strongly on line tension,
area-to-volume ratio.    
The early dynamics in this case, is governed by the coalescence of 
small flat patches. In the presence of excess area,
the later dynamics dynamics is characterized by the coalescence of caps.
The crossover between the two regimes depends strongly on line tension, 
and includes an intermediate vesiculation regime for high enough line tension.
In the case of tensed vesicles, no crossover is observed in the dynamics.
In the case of critical quenches, the growth dynamics is qualitatively different
and no crossovers were observed.

\section*{Acknowledgments}
The authors would like to thank O.G. Mouritsen and L. Bagatolli, M. Rao and G.I. Menon for useful
discussions and critical comments. MEMPHYS is supported by the Danish National Research Foundation. Part of
the simulations were carried out at the Danish Center for Scientific Computing.
This work is supported in part by the Petroleum Research Fund of the American Chemical Society.

\section*{Appendix A: Numerical Extraction of the Bending Modulus}

The bending modulus is extracted from the power spectrum of the long-wavelength out-of-plane fluctuations of an open
lipid membrane, extending along the $xy$-plane, in a box with periodic boundary conditions along the three directions. 
The Helfrich Hamiltonian of a membrane, in terms of the principal curvatures, $c_1(\rr)$ and
$c_2(\rr)$, is given by
\BE\label{eq:helfrich}
{\cal H}({c_1},{c_2})=\int^{\cal A} da\left[\sigma+\frac{\kappa}{2}(c_1+c_2-2c_0)^2+\overline{\kappa}{c_1c_2}+\cdots\right]
\EE
where $\sigma$, $\kappa$ and $\overline\kappa$ are the tension, the bending modulus and the Gaussian bending modulus of the
membrane respectively. $c_0$ is its spontaneous curvature. In the case of a one-component membrane, and if the lipid lateral densities
of the two leaflets are equal, the spontaneous curvature vanishes. Furthermore, if the membrane conserves its topology, the integral of the Gaussian term is independent of the membrane conformation. If we assume that the height of the membrane is
represented by a single-valued function $h({\bf x})$, and in the case of small fluctuations, the Hamiltonian can be rewritten
in the Monge representation as,
\BE
{\cal H}({h})=\int d{\bf x}^{L^2}=\left[\frac{\sigma}{2}(\nabla_{\bf x} h)^2+\frac{\kappa}{2}(\nabla_{\bf x}^2 h)^2+
\cdots\right].
\EE
The structure of the membrane can then be inferred from the structure factor, defined as the Fourier transform of
the height-height correlation function,
\BE
S_h(\qq)=\frac{1}{L^2} \langle\mid {\tilde h}_\qq \mid^2\rangle=
\biggl{\langle} \mid \int^{L^2}d\xx e^{-i\qq\cdot\xx} h(\xx)
\mid^2 \biggl{\rangle},
\EE
where $\qq=(q_x,q_y)$. If the higher order powers of $h$ are omitted in the Hamiltonian, and after the equi-partition theorem is invoked, one finds the the membrane height structure factor is given by
\BE
S_h(\qq)=\frac{2k_BT}{\sigma q^2+\kappa q^4+{\cal O}(q^6)}.
\EE
The membrane position, $h(\xx)$, is defined as the location of the median point of the hydrophobic region. In Fig.~14, the 
structure factor for a one component open membrane in a box of dimensions $L_x=L_y=86r_c$ and $L_z=40 r_c$ is shown. 
The other parameters are exactly the same as those presented in Sec. II.
The bending modulus and the tension on the membrane are obtained from 
examining the structure factor at small wavevectors. The 
tension on the membrane is found from the intercept of the $\sigma(q)$ {\em vs.} $q^2$ curve with the $y$-axis where, 
\BE
\sigma(q)=2 k_BT/q^2S_h(q).
\EE
The bending modulus is
obtained from the slope of $\sigma(q)$, at small $q$'s, as illustrated in Fig.~14.
We obtained a value $\kappa\approx 8\epsilon$, when $k_BT=\epsilon$. The value of $\kappa$ in our model is reasonably in 
agreement with the experimental values for lipid bilayers. 

\begin{figure}
\includegraphics[scale=0.5]{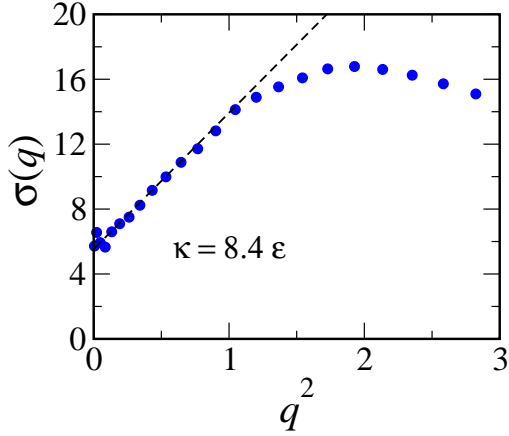}
\caption{$\sigma(q)=2/q^2S_h(q)$ versus $q^2$
obtained from a simulation of a one-component open membrane with projected
area $L_x\times L_y=86r_c\times 86r_c$. The slope of the curve at small 
wavevectors determines the bending modulus, and the intercept with the $y$-axis
determines the surface tension of the membrane.}
\label{bendmod}
\end{figure}

\section*{Appendix B: Numerical Extraction of the Line Tension}

The in-plane tension of a membrane extending along the $xy$-plane is calculated from the pressure tensor as
\BE\label{eq:sigma}
\sigma= L_z\left[P_{zz}-\frac{1}{2}\left(P_{xx}+P_{yy}\right)\right],
\EE
where the pressure tensor ${\bf P}_{\alpha\beta}$ is calculated using the Irving-Kirkwood formalism~\cite{irving50},
\BE
{\bf P}=\frac{1}{\Delta V}\sum_{i}\left\langle m_i \vv_i\vv_i+
\frac{1}{2}\sum_{i}\sum_{j\ne i}\rr_{ji}{\bf F}^{(C)}_{ji}\right\rangle.
\EE

To calculate the line tension, the membrane is prepared such that it consists of {\em A} and {\em B} coexisting phases,
separated by two interfaces which are parallel to the $x$-axis. The tension of the membrane now contains a 
two-dimensional bulk component plus a one-dimensional contribution due to the interfaces between the 
{\em A} and {\em B} components. The line tension is then calculated from the difference between the tension of a membrane
with two interfaces between the {\em A} and {\em B} components and that of a membrane composed with $A$-lipids only,
\BE
\lambda=\frac{L_y}{2}\left(\sigma'-\sigma\right).
\EE

$\sigma$ and $\sigma'$ were calculated on system sizes with dimensions $L_x=L_y=86r_c$ and $L_z=40r_c$.
We found that $\lambda\approx 7.4\epsilon/r_c$ and $5.5\epsilon/r_c$ for
$a_{h_1h_2}=50\epsilon$ and $100\epsilon$, respectively. 
If we assume that the lipid bilayer thickness is 4 nm, then our 
numerically calculated line tensions correspond to $2.3\times 10^{-17} {\rm J}\mu{\rm m}$ and
$1.73\times 10^{-17} {\rm J}\mu{\rm m}$, respectively. 

Although there are no experimental data for the line tension between coexisting lipid phases, a simple estimation 
can be given by $\lambda \approx\Delta U/{l}$, where $\Delta U=(U_{AA}-U_{BB})/2-U_{AB}\sim 10 k_BT$,
where $U_{\alpha\beta}$ are the various lipid pair interactions, 
and $l\approx 0.8$ nm is the lateral length scale associated with a lipid molecule. One then finds 
$\lambda \sim 10^{-17} {\rm J}\mu{\rm m}$, in agreement with our numerical values~\cite{sackman95}.


\begin{thebibliography}{00}
\bibitem{footnote1} The weak dependence of the surface tension on $\Delta t$ 
~\cite{ask05} does not affect the generic results of the present study.
\end{thebibliography}
\end{document}